 \documentclass[twocolumn]{aastex6}
 \usepackage{amsmath}

\fullcollaborationName{The Friends of AASTeX Collaboration}

\begin{document}


\title{A mixture evolution scenario of the AGN radio luminosity function. II. Do low- and high-power radio-loud AGNs evolve differently?}



\author{Zunli Yuan\altaffilmark{1,2,3}, Jiancheng Wang\altaffilmark{1,2,3}, Ming Zhou\altaffilmark{1,2,3} Longhua Qin\altaffilmark{1,2,3,4} and Jirong Mao\altaffilmark{1,2,3}}
\affil{yuanzunli@ynao.ac.cn}







\altaffiltext{1}{Yunnan Observatories, Chinese Academy of Sciences, Kunming 650216, P. R. China}
\altaffiltext{2}{Center for Astronomical Mega-Science, Chinese Academy of Sciences, Beijing 100012, P. R. China}
\altaffiltext{3}{Key Laboratory for the Structure and Evolution of Celestial Objects, Chinese Academy of Sciences, Kunming 650216, P. R. China}
\altaffiltext{4}{University of Chinese Academy of Sciences, Beijing 100049, P. R. China}

\begin{abstract}

Following previous work, we further confirm that the cosmic evolution of steep-spectrum radio-loud AGNs (active galactic nuclei) can be reproduced by a simple combination of density evolution (DE) and luminosity evolution (LE). This mixture evolution scenario can naturally explain the luminosity-dependent evolution of radio-loud AGNs. Our models successfully fitted a large amount of data on radio luminosity functions (RLFs) of steep-spectrum sources and multi-frequency source counts. The modeling indicates that the DE slowly increase as $(1+z)^{0.3 \thicksim 1.3}$ out to $z \thicksim 0.8$, and then rapidly decreases as $(1+z)^{-6.8 \thicksim -5.7}$, while the LE rapidly increase as $(1+z)^{4.8}$ out to a higher redshift (at least $z>3.5$). We find a high-redshift decline (i.e. redshift cutoff) in the number density of steep-spectrum radio sources, but we cannot conclude whether such decline is sharp or shallow. We believe that whether a redshift cutoff occurs or not depends mainly on DE, while its steepness is decided by LE, which, however, cannot be well constrained due to the lack of high-redshift samples. Most intriguingly, according to our mixture evolution scenario, there appears to be no need for different evolution for the low- and high-power radio-loud AGNs. Both types of sources experience the same combined evolution of DE and LE.

\end{abstract}





\section{Introduction}
The luminosity function (LF) has always been an important and common tool for understanding the evolution of AGNs over cosmic time. The radio luminosity function (RLF) is particularly useful for exploring the the space density of AGNs at high redshift because the radio emission has the advantage of being free from dust obscuration \citep[e.g.,][]{2007MNRAS.375.1349C,2015MNRAS.449.2818T}. Over the past decades, it has been well known that AGNs evolve strongly, i.e., their LF changes with redshift, with their luminosities and/or their numbers being different from what they are at $z \thicksim 0$ \citep{2016A&ARv..24...13P}. Obviously, there are at least two physically different evolution mechanisms. The first one is density evolution (DE) whose physical meaning is whether the sources are more or less numerous than that of today. The second one is luminosity evolution (LE) which represents whether the sources are more or less luminous than that of today \citep[][hereafter Paper I]{2016ApJ...820...65Y}. We can easily know if a sample of objects evolves or not by their LF, but it is difficult to distinguish whether such an evolution is purely due to DE or due to LE \citep[e.g.,][]{2012MNRAS.422.2274C}. As the two evolution mechanisms are degenerated \citep[e.g.,][]{2009ApJ...696...24S}. Therefore, the nature of evolution is still an open question \citep{2010MNRAS.404..532M}.

After the influential work of \citet{1990MNRAS.247...19D}, there have been many papers that discussed the luminosity-dependent evolution of radio-loud AGNs \citep[e.g.,][]{2001MNRAS.328..882W,2004MNRAS.352..909C,2011MNRAS.416.1900R,2013MNRAS.436.1084M,2017A&A...602A...6S}. The key point is that powerful radio-loud AGNs undergo very rapid evolution to a redshift of $z \thicksim 3$ , while their lower luminosity counterparts only experience much milder positive evolution to $z \thicksim 1$. In other words, the position of the RLF peak is luminosity dependent. This can be interpreted as a sign of cosmic downsizing, where the most massive black holes form at earlier epochs than their less massive counterparts \citep{2015A&A...581A..96R}. However, as pointed by \citet{2016A&ARv..24...13P}, we need to be careful about how to interpret the luminosity-dependent evolution phenomenon. Indeed, in Paper I, we have shown that the apparently complex behavior displayed by the steep-spectrum radio sources studied by \citet[][hereafter R15]{2015A&A...581A..96R} can be easily reproduced by a simple combination of DE and LE. In such a mixture evolution scenario, the luminosity-dependent evolution is naturally present.

In addition, the effect of spectral index distribution was probably underestimated in the general study of RLF. Although \citet{2000MNRAS.319..121J} highlighted the importance of a distribution in spectral index in the parametric modeling of RLF \citep[also see][]{2001MNRAS.327..907J}, most works on RLFs did not fully consider this effect, as they often use bivariate RLF estimators. Recently, we argued that for the previous results based on bivariate RLF estimators, which did not sufficiently consider the effect of spectral index distribution, the significance/degree of luminosity-dependent evolution can be magnified \citep[][and references therein]{2016ApJ...829...95Y}. In this work, we will use a trivariate RLF estimator, that incorporates the spectral index distribution, to further develop the mixture evolution scenario.

Throughout the paper, we adopt a Lambda Cold Dark Matter cosmology with the parameters $\Omega_{m}$ = 0.27,  $\Omega_{\Lambda}$ = 0.73, and $H_{0}$ = 71 km s$^{-1}$ Mpc$^{-1}$.

\section{Methods}
The data used in this work is the same as these in Paper I, i.e., the combined sample established by \citet{2012ApJ...744...84Y}, which consists of four subsamples: the MRC1 \citep{1996ApJS..107...19M}, the MS4 \citep{2006AJ....131..100B}, the BRL \citep{1999MNRAS.310..223B} and the 3CRR \citep{1983MNRAS.204..151L} samples. The sources in our sample are all steep-spectrum, consisting mainly of radio galaxies (RGs) and steep-spectrum quasars. Although large, a disadvantage of our combined sample is that it lacks faint radio sources. The deepest sub-sample is the MRC1 sample whose flux limit is 0.95 Jy at 408 MHz. Therefore, we use the 408 MHz radio counts and 1.4 GHz local RLF to provide additional constraints for the modeling process. In addition, we use a trivariate RLF estimator in which the spectral index distribution is incorporated to give a more accurate analysis of the RLF.

\subsection{The trivariate RLF}

Following \citet{2016ApJ...829...95Y}, we define a trivariate RLF as the number of sources per comoving volume $V(z)$ with radio luminosities in the range $L,L+dL$, and with spectral indexes in the range $\alpha,\alpha+d\alpha$:
\begin{eqnarray}
\label{aaa}
\Phi(\alpha,z,L)=\frac{d^{3}N}{d\alpha dz dL}.
\end{eqnarray}
If the spectral index is independent of redshift and luminosity, $\Phi(\alpha,z,L)$ can be written as
\begin{eqnarray}
\label{aaa1}
\frac{d^{2}N}{dL dV}\times\frac{dN}{d\alpha}\times\frac{dV}{dz}=\rho(z,L)\times\frac{dN}{d\alpha}\times\frac{dV}{dz}.
\end{eqnarray}
where $\rho(z,L)$ is the common defined RLF (or referred as bivariate RLF), and $dV/dz$ is the comoving volume element. The function $dN/d\alpha$ is the intrinsic spectral index distribution.
$\Phi(\alpha,z,L)$ is related to the probability distribution of $(\alpha,z,L)$ by
\begin{eqnarray}
\label{eqpc}
p(\alpha,z,L)=\frac{1}{N_{tot}}\Phi(\alpha,z,L)\frac{dV}{dz}.
\end{eqnarray}
where $N_{tot}$ is the total number of sources in unit solid angle in the universe, and is given by the integral of $\Phi$ over $\alpha$, $L$ and $V(z)$. The likelihood function $p(\alpha_{obs},z_{obs},L_{obs}|\theta)$ for the observed data can be derived, once we assume a parametric form for $\Phi(\alpha,z,L)$, with parameters $\theta$. \citet{1983ApJ...269...35M} give a likelihood function based on the Poisson distribution, and for the three-dimensional case, their original definition is updated as $S=-2ln(p(\alpha_{obs},z_{obs},L_{obs}|\theta))$ \citep[also see][]{2001MNRAS.327..907J,2016ApJ...829...95Y}. Dropping the terms independent of the model parameters, we have
\begin{eqnarray}
\label{likelihood2}
\begin{aligned}
S=-2\sum_{i}^{N_{obs}}&ln[\Phi(\alpha_i,z_{i},L_{i})]+\\
&2\int\int\int\Phi(\alpha,z,L)\Omega(\alpha,z,L)\frac{dV}{dz}d\alpha dz dL.
\end{aligned}
\end{eqnarray}
Considering the limits of the integral in $S'$, we have
\begin{eqnarray}
\label{likelihood3}
\begin{aligned}
S&=-2\sum_{i}^{N_{obs}}ln[\Phi(\alpha_i,z_{i},L_{i})]+\\
&2\sum_{j}^{4}\Omega^j \int^{\alpha_2^j}_{\alpha_1^j}d\alpha \int^{z_{2}^j}_{z_{1}^j}dz\frac{dV}{dz}\int_{max[L_1^j,L^j_{lim}(\alpha,z)]}^{L_2^j}\Phi(\alpha,z,L)dL,
\end{aligned}
\end{eqnarray}
where $(\alpha_1^j,\alpha_2^j)$, $(z_1^j,z_2^j)$ and $(L_1^j,L_2^j)$ are the spectral index, redshift and luminosity limits of the $j$th subsample respectively, $L_{lim}^j(\alpha,z)$ is the luminosity limit surface corresponding to the flux density limit, and $\Omega^j$ is the solid angle subtended by the subsample $j$.

Following \citet{2001MNRAS.322..536W}, we use the source counts and local RLF to provide additional constraints for the fitting process. The source counts and local RLF are one-dimensional functions, and their $\chi^2$ is evaluated as
\begin{eqnarray}
\label{chi2}
\chi^2=\sum_{i=1}^{N}(\frac{f_{data~i}-f_{mod~i}}{\sigma_{data~i}})^2,
\end{eqnarray}
which is related to the likelihood by $\chi^2=-2$ln(likelihood), i.e. the same form as $S$. Therefore, we can define a new function $S_{all}$ \citep[also see][]{2001MNRAS.322..536W} that combines the constraints from all three types of data:
\begin{eqnarray}
\label{chi2all}
S_{all}=\chi^2_{SC} + \chi^2_{LRLF} + S,
\end{eqnarray}
where $\chi^2_{SC}$ and $\chi^2_{LRLF}$ represent the values of $\chi^2$ for the source counts and local RLF, respectively. We can obtain the best estimates for the model parameters by minimizing $S_{all}$. In this work, we use a Bayesian method as described in Paper I to obtain the best estimates for the model parameters and their probability distribution \citep[also see][]{2002PhRvD..66j3511L}.

\subsection{Radio source counts}

The source counts is a classical statistical tool to provide useful information, as its shape is tightly related to the evolutionary properties of the sources and also to the geometry of the universe \citep{2016A&ARv..24...13P}. The counts are usually presented in differential form $dN/dS$, giving the number of sources per flux density per steradian. Usually, multiplied by $S^{2.5}$, $dN/dS$ is normalized to the `Euclidean' form \citep[e.g.,][]{2010A&ARv..18....1D}. If we know the RLF of the sources, we can derive the differential counts by the following equation.
\begin{eqnarray}
\label{aaa}
\frac{dN}{dS}=4\pi\frac{c}{H_{0}}\int_{z_{1}}^{z_{2}}\int_{\alpha_{1}}^{\alpha_{2}}\frac{\Phi(\alpha,z,L(\alpha,z,S))D^{4}_{L}(z)d\alpha dz}{(1+z)^{(3-\alpha)}\sqrt{\Omega_M(1+z)^3+\Omega_{\Lambda}}},
\end{eqnarray}
where $c$ is the speed of light, $\Phi(\alpha,z,L)$ is the RLF, $D_{L}(z)$ is the luminosity distance, $z_1$ and $z_2$, $\alpha_1$ and $\alpha_2$ represent the range of integration in redshift and spectral index \citep[e.g.,][]{2016A&ARv..24...13P}.

\citet{2010A&ARv..18....1D} presented a compilation of radio source counts at multiple frequencies \citep[also see][]{2010MNRAS.404..532M}. In this work, we use the 408 MHz counts to provide an additional constraint for the model, and also use the 1.4 GHz counts to compare with our model prediction.

\subsection{The local RLF}
Thanks to the combined use of large radio surveys, such as NVSS (NRAO VLA Sky Survey) and FIRST (Faint Images of the Radio Sky at Twenty centimeters), and large-area spectroscopic surveys, the local RLF at 1.4 GHz has been well determined \citep[e.g.,][]{2000A&A...360..463M,2002AJ....124..675C,2002MNRAS.333..100M,2002MNRAS.329..227S,2005MNRAS.362....9B,2007MNRAS.375..931M}. We mainly refer to the local RLF of \citet{2007MNRAS.375..931M} to constrain our model. Since their local RLF is not at the same frequency as that of this work, we have shifted luminosities from 1.4 GHz to 408 MHz according to a power law ($L/\nu\propto \nu^{-\alpha}$, with $\alpha$ = 0.75).

\subsection{The form of the RLF}
\label{form}

Without loss of generality, we can write an RLF as
\begin{eqnarray}
\label{aaa}
\Phi(\alpha,z,L)=e_1(z)\rho(z=0,L/e_2(z),\eta^j)\frac{dN}{d\alpha},
\end{eqnarray}
where $e_1(z)$ and $e_2(z)$ describes, respectively the DE and LE functions with redshift, and $\eta^j$ stands for parameters that describe the shape of the RLF. If the parameters $\eta^j$ are constant, this means that the RLF shape is preserved. Otherwise, if the the parameters $\eta^j$ have some redshift dependence, this is equivalent to having luminosity-dependent density evolution \citep[for details, see][]{2013ApJ...764...43S,2014ApJ...786..109S}. In our mixture evolution model, a non-evolving shape for the LF is assumed (i.e., $\eta^j$  = const, independent of L and z). Since the DE function $e_1(z)$ in Paper I is not equal to one at $z=0$, here we use a new form for $e_1(z)$:
\begin{eqnarray}
\label{f_d}
e_1(z)=p_0\left[ \left(\frac{1+z_c}{1+z}\right)^{p_1} + \left(\frac{1+z_c}{1+z}\right)^{p_2} \right]^{-1},
\end{eqnarray}
where $p_1$, $p_2$ and $z_c$ are free parameters, while $p_0$ is the normalized parameter of $e_1(z)$ given by
\begin{eqnarray}
\label{f_d}
p_0=\left[ (1+z_c)^{p_1} + (1+z_c)^{p_2} \right].
\end{eqnarray}

Traditionally, the local RLF $\rho(z=0,L/e_2(z=0))$ is often described by a double-power-law form \citep[e.g.,][]{2012ApJ...751..108A,2000MNRAS.317.1014B}. Enlightened by \citet{1995ApJ...438..623P} and \citet{2007ApJ...654..731H}, we use a modified Schechter function to describe the local RLF, given by
\begin{eqnarray}
\label{eqn:LRLF}
\begin{aligned}
\rho(z=0,L/e_2&(z=0))=\frac{dN}{dL} \\
&=\frac{\phi}{\ln(10)L}\left( \frac{L}{L_*} \right)^{-\beta} \exp\left[-\left( \frac{L}{L_{*}}\right)^{\gamma}\right],
\end{aligned}
\end{eqnarray}
where $\phi$ is the normalization factor. The value of $\phi$ depends on whether a trivariate RLF, $\Phi(\alpha,z,L)$, or bivariate RLF, $\rho(z,L)$, is considered, where $\phi$ equals $\phi_0$ or $\phi_1$, respectively. $\phi_1$ and $\phi_0$ are not independent, and they are related by
\begin{eqnarray}
\label{eeeeee}
\int\int\rho(z,L)\frac{dV}{dz}dzdL\equiv \int\int\int\Phi(\alpha,z,L)\frac{dV}{dz}d\alpha dzdL.
\end{eqnarray}

The LE function $e_2(z)$ has three different forms depending upon the model: the traditional power-law form
\begin{eqnarray}
e_2(z)=(1+z)^{k_1}
\end{eqnarray}
for model A, a modified power-law form
\begin{eqnarray}
e_2(z)=(1+z)^{k_1}e^{-z^2/2k_2^2}
\label{LEB}
\end{eqnarray}
for model B, and the polynomial form used by \citet{1990MNRAS.247...19D} and \citet{2000MNRAS.317.1014B}
\begin{eqnarray}
\label{eqn:flc}
e_2(z)=10^{k_1z^2+k_2z}
\label{LEC}
\end{eqnarray}
for model C.

\begin{table*}[!t]
\tablewidth{0pt}
\renewcommand{\arraystretch}{1.5}
\caption{Best-fit Parameters}
\begin{center}
\setlength{\tabcolsep}{1mm}{
\begin{tabular}{lccccccccccl}
\hline\hline

\colhead{Model~~$\log_{10}\phi_0$} & \colhead{$\log_{10}\phi_1$} &
\colhead{$\log_{10}L_*$}  & \colhead{$\beta$} &
\colhead{$\gamma$}        & \colhead{z$_{\rm c}$} &
\colhead{p$_{\rm 1}$}     & \colhead{p$_{\rm 2}$} &
\colhead{k$_{\rm 1}$}     & \colhead{k$_{\rm 2}$} &
\colhead{$\mu$}           & \colhead{$\sigma$} \\
\hline

~A~~~ -4.63$_{-0.12}^{+0.13}$ & -4.85$_{-0.12}^{+0.13}$ & 24.68$_{-0.17}^{+0.16}$ & 0.44$_{-0.02}^{+0.02}$ & 0.31$_{-0.01}^{+ 0.01}$ & 0.86$_{-0.09}^{+0.10}$ & 0.31$_{-0.26}^{+0.22}$ & -5.92$_{-0.39}^{+0.18}$ & 4.73$_{-0.09}^{+0.16}$ & ... & -0.09$_{-0.01}^{+0.00}$ & 0.24$_{-0.01}^{+0.00}$ \\

~B~~~ -4.61$_{-0.15}^{+0.10}$ & -4.84$_{-0.15}^{+0.10}$ & 24.66$_{-0.14}^{+0.19}$ & 0.44$_{-0.02}^{+0.02}$ & 0.31$_{-0.01}^{+ 0.02}$ & 0.82$_{-0.09}^{+0.10}$ & 0.28$_{-0.25}^{+0.30}$ & -5.67$_{-0.34}^{+0.38}$ & 4.85$_{-0.13}^{+0.14}$ & 4.64$_{-1.20}^{+1.46}$ & -0.09$_{-0.01}^{+0.01}$ & 0.24$_{-0.00}^{+0.00}$ \\

~C~~~ -4.72$_{-0.35}^{+0.43}$ & -4.95$_{-0.35}^{+0.43}$ & 24.79$_{-0.59}^{+0.42}$ & 0.45$_{-0.07}^{+0.06}$ & 0.31$_{-0.04}^{+ 0.03}$ & 0.78$_{-0.17}^{+0.21}$ & 1.29$_{-0.65}^{+0.78}$ & -6.80$_{-1.46}^{+1.34}$ & -0.16$_{-0.06}^{+0.05}$ & 1.44$_{-0.17}^{+0.17}$ & -0.09$_{-0.02}^{+0.02}$ & 0.24$_{-0.01}^{+0.02}$ \\

\hline
\end{tabular}}
\end{center}
~~Units -- $\phi_0$ and $\phi_1$: [${\rm Mpc^{-3}}$],\,\, $L_*$: [${\rm W Hz^{-1}}$]. The best-fitting parameters as well as their 1 $\sigma$ errors for model A, B and C.
\label{tab:fit}
\end{table*}

For the intrinsic spectral index distribution $dN/d\alpha$, \citet{2012MNRAS.422.2274C} found that the Gaussian forms describe well their shape for both steep-($\alpha>0.5$) and flat-spectrum ($\alpha<0.5$) sources. Here we give a new consideration: the typical value of $\alpha$ for steep-spectrum sources is 0.75, and its lower limit is 0.5 (according to the definition of ``steep-spectrum'') while its upper limit can reach 2.0 or even higher. Obviously, a right-skewed distribution could be more reasonable for describing the intrinsic spectral index distribution. Here we use a logarithm-normal function to model $dN/d\alpha$:

\begin{eqnarray}
\label{SID}
\frac{dN}{d\alpha}=\frac{1}{\alpha}e^{-\frac{(\ln\alpha-\mu)^2}{2\sigma^2}}
\end{eqnarray}

\begin{figure*}[!htb]
\centering
\includegraphics[height=7cm,width=8cm]{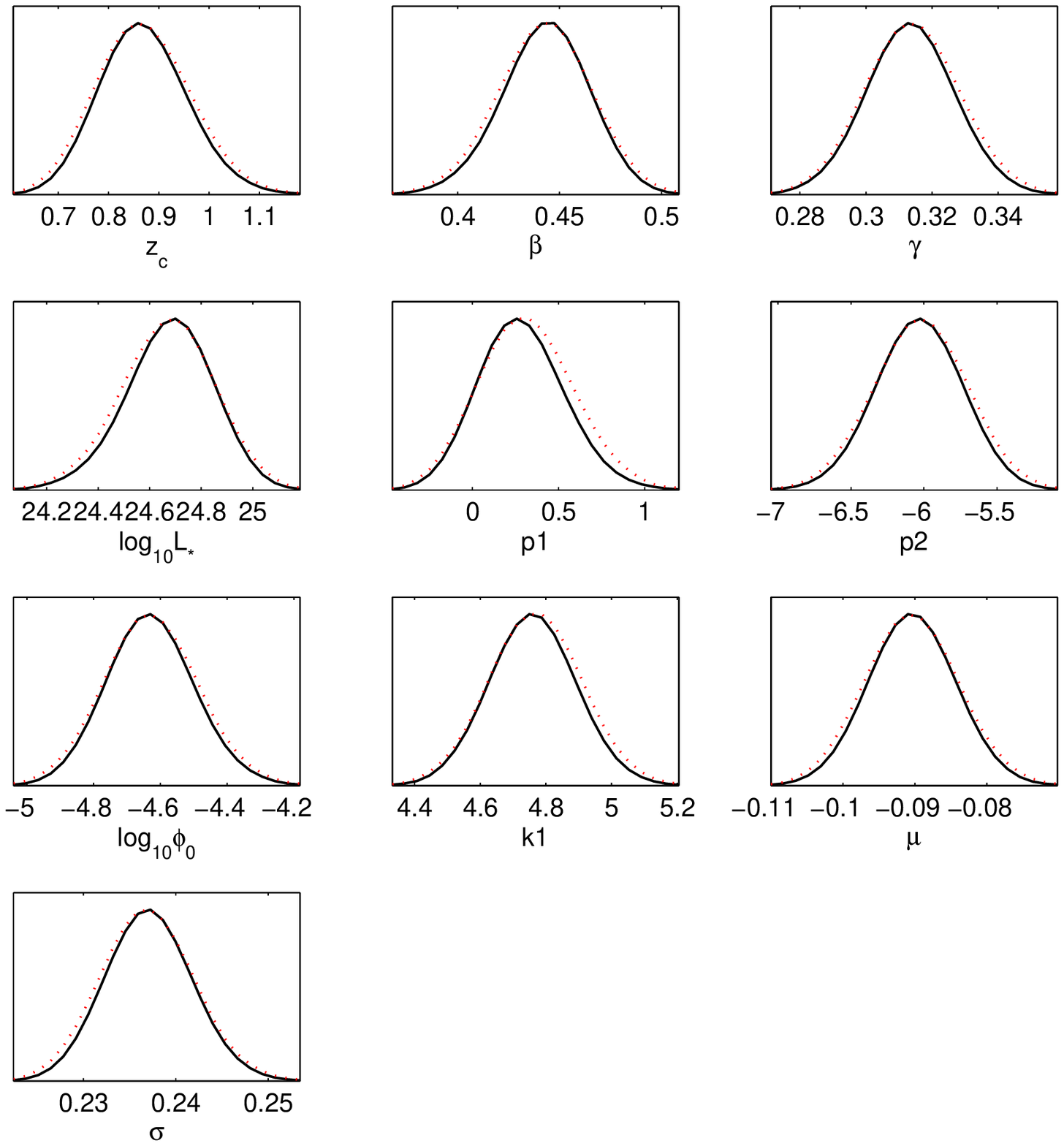}
\includegraphics[height=7cm,width=8cm]{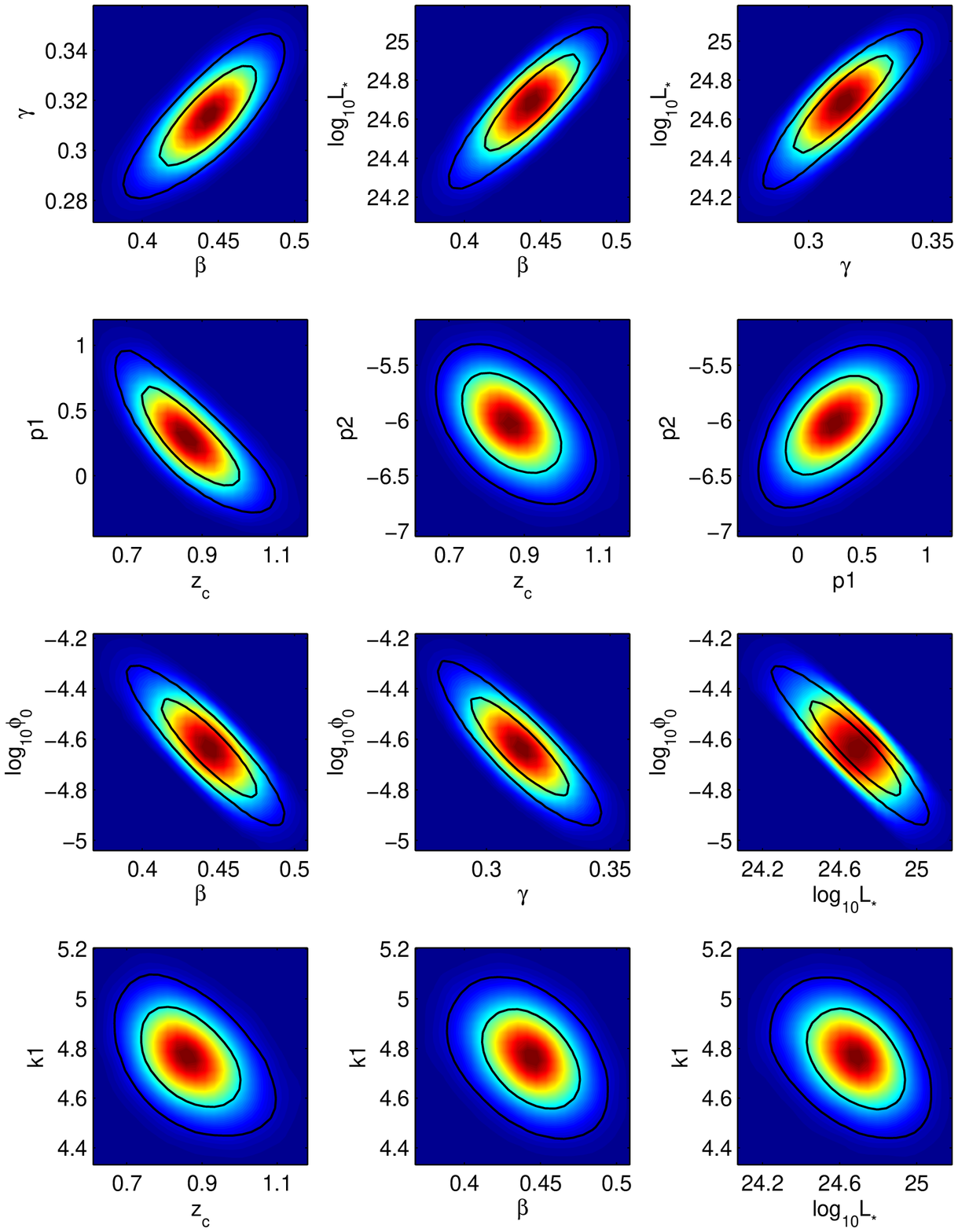}
\\[18pt]
\includegraphics[height=7cm,width=8cm]{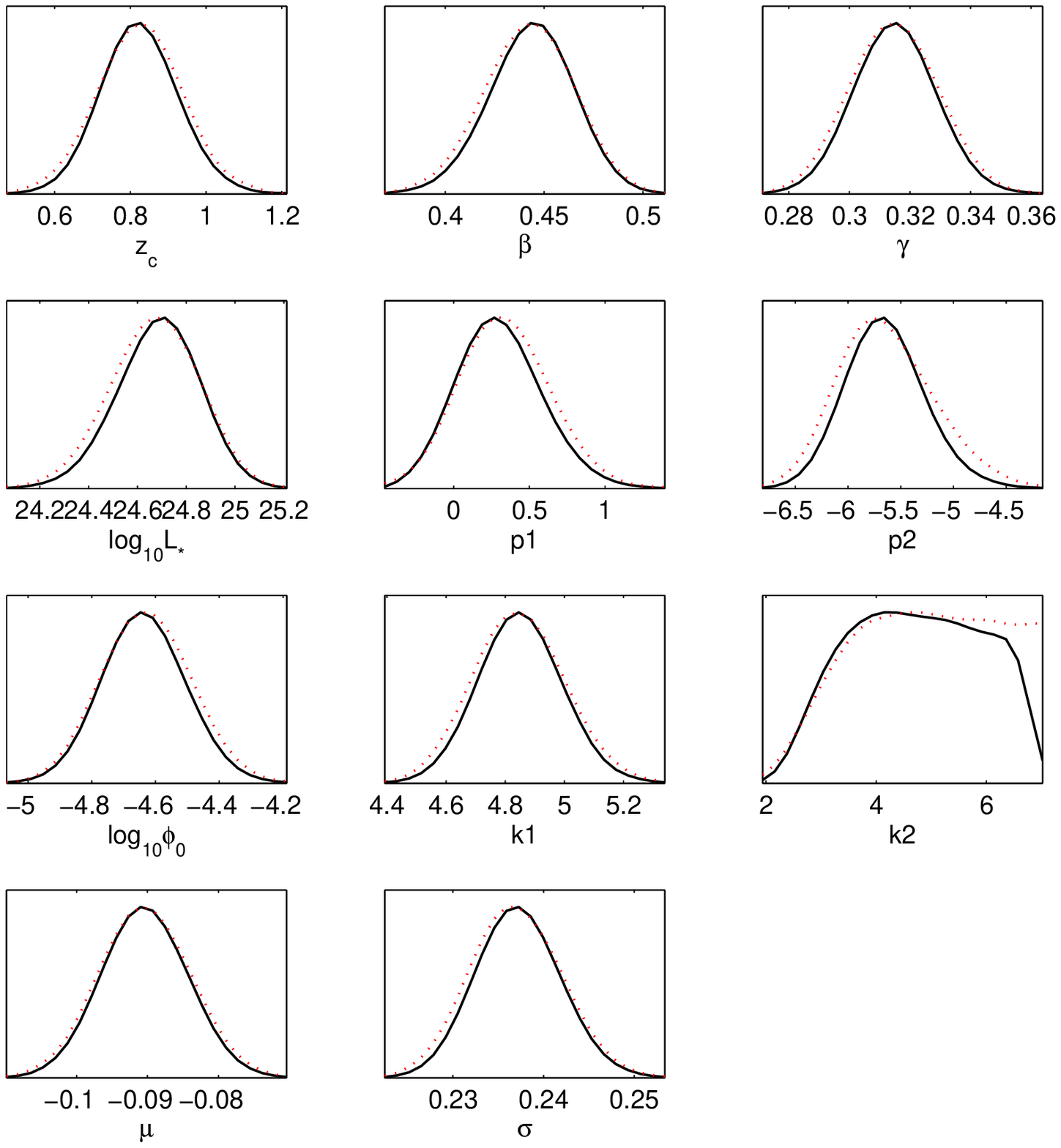}
\includegraphics[height=7cm,width=8cm]{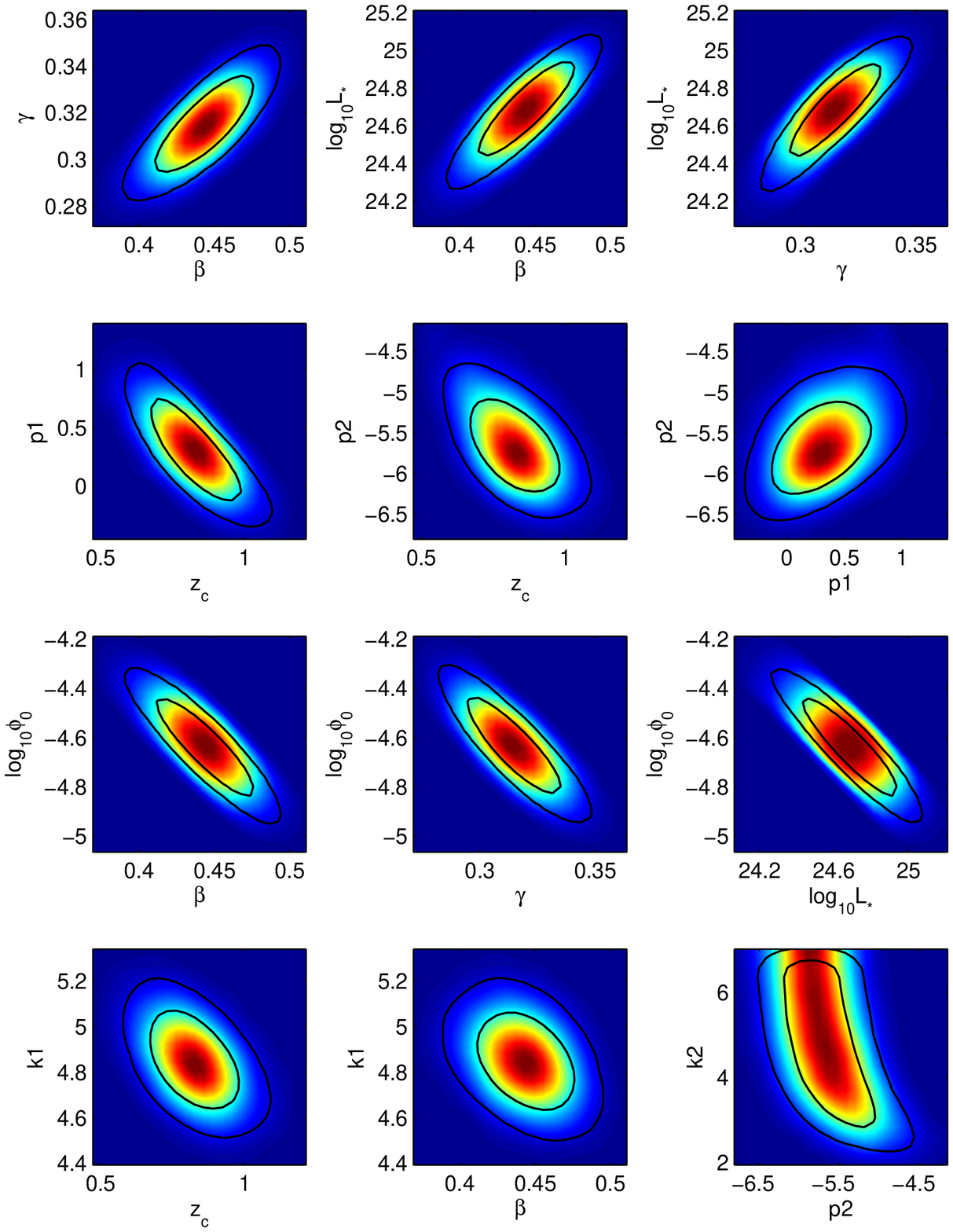}
\\[18pt]
\includegraphics[height=7cm,width=8cm]{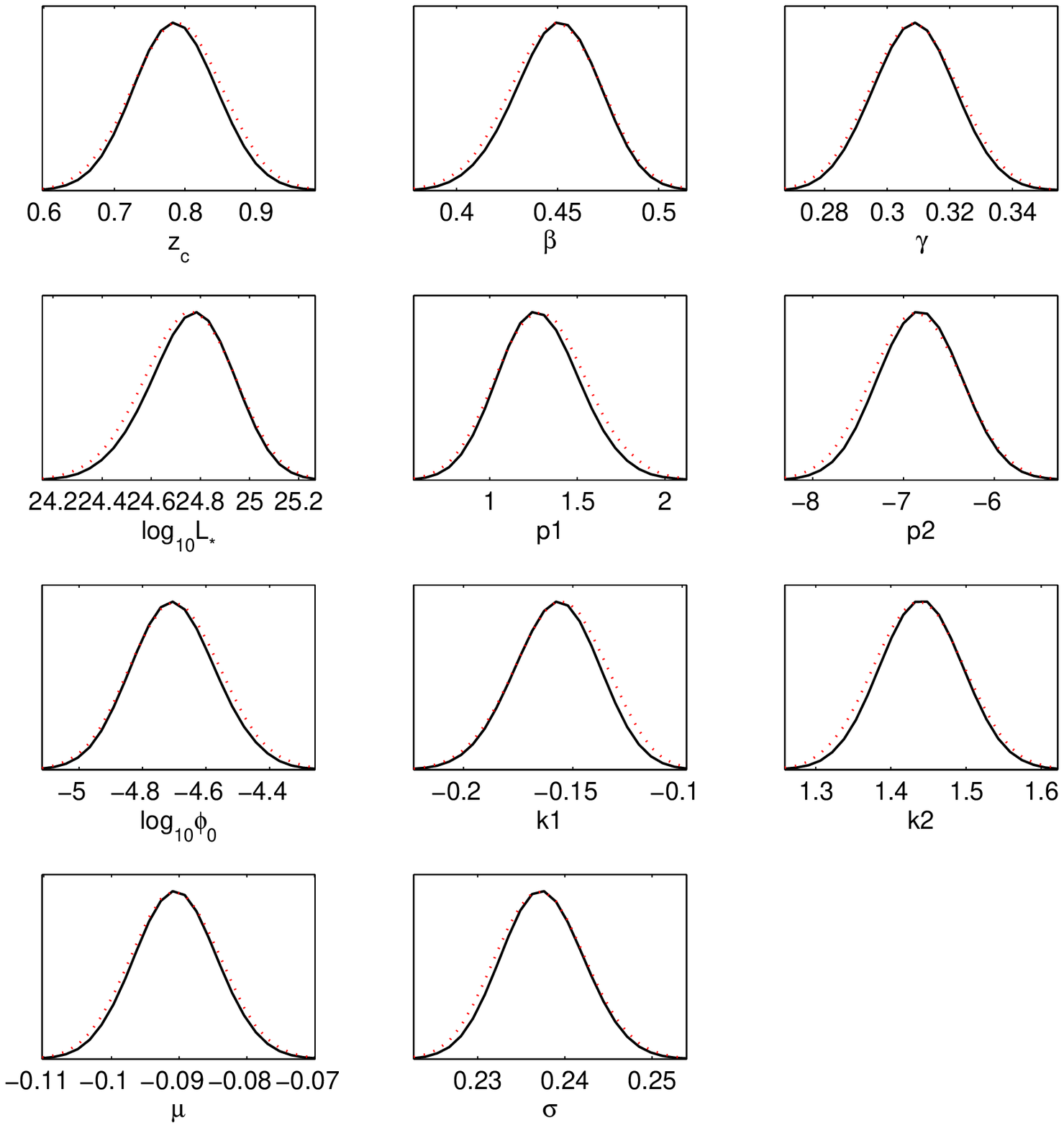}
\includegraphics[height=7cm,width=8cm]{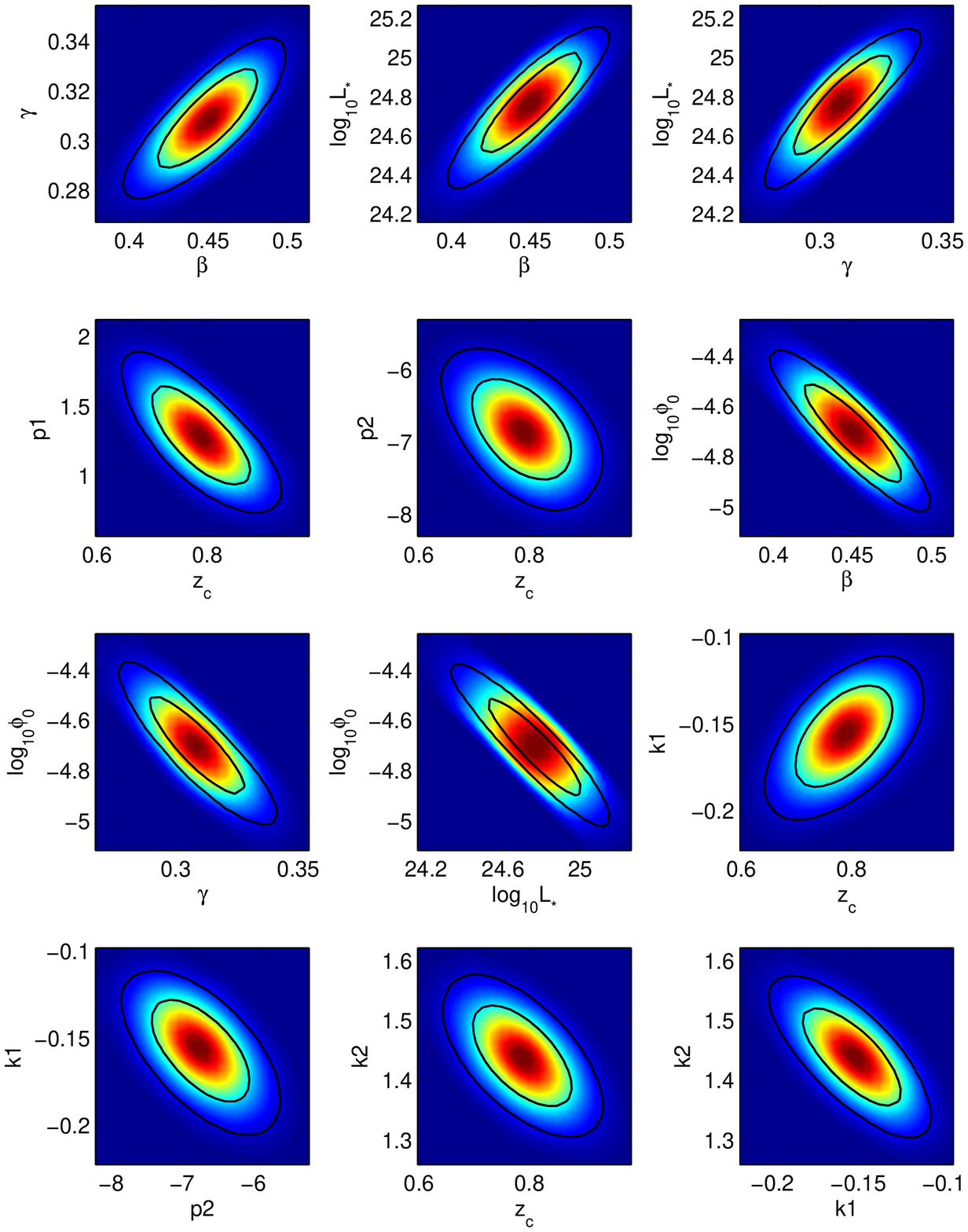}
\caption{Left: 1D probability distribution of the parameters in the models \citep[the red dash-dotted curves are the mean likelihoods of MCMC samples and the black solid curves are the marginalized probabilities, see][]{2002PhRvD..66j3511L}; Right: 2D confidence contours of the parameters. The contours are for 1 and 2 $\sigma$ levels. For the 2D confidence regions of the parameters, only combinations with relatively large correlations are presented. The upper, middle and lower panels are for models A, B, and C respectively.}
\label{fig_mcmc}
\end{figure*}

\begin{figure}[!h]
\centering
\includegraphics[width=1.0\columnwidth]{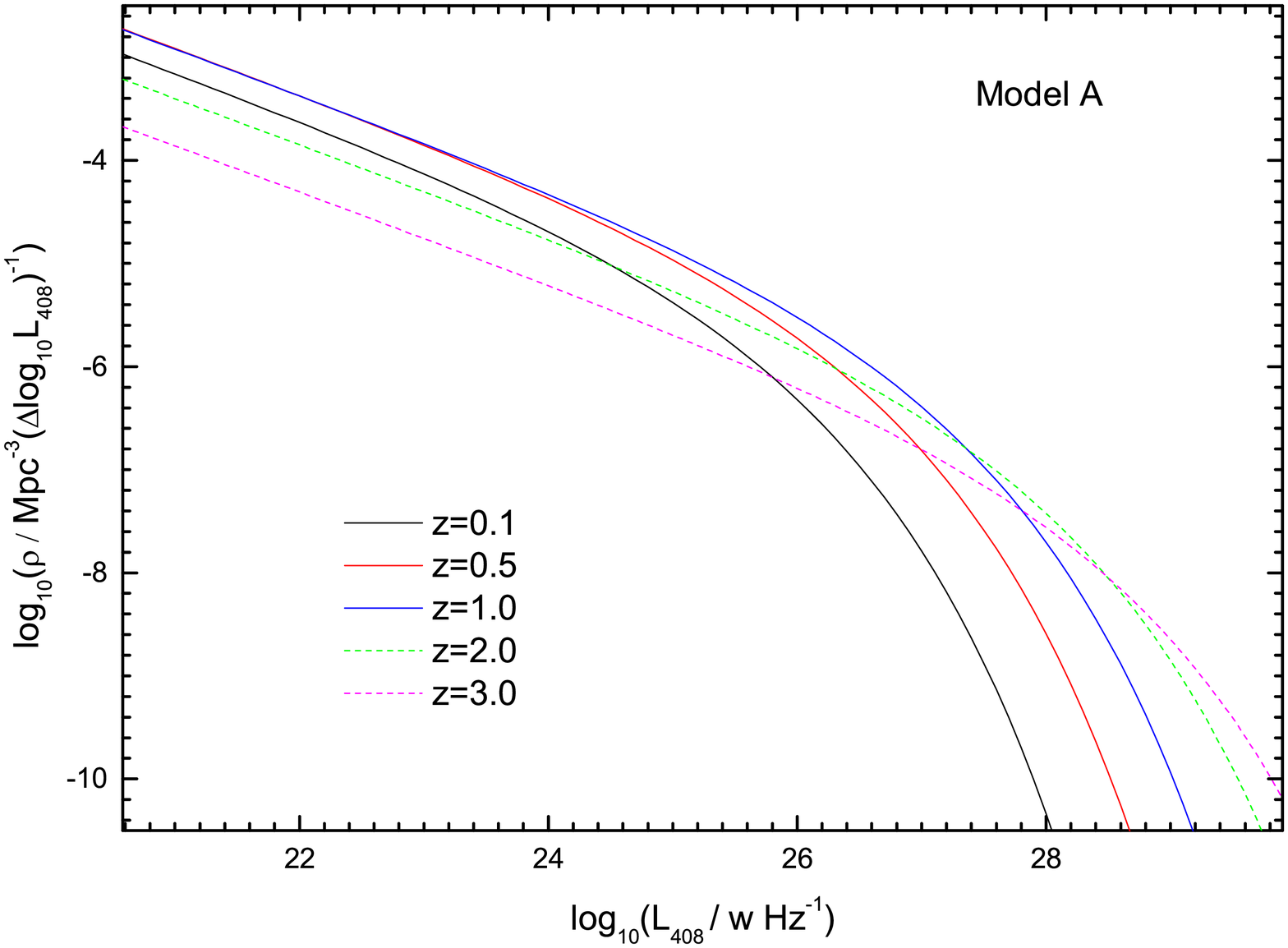}
\caption{The 408 MHz RLF derived for model A at z=0.1, 0.5, 1.0, 2.0, and 3.0 (black, red, blue solid lines, green and violet dashed lines respectively).}
\label{RLF-L}
\end{figure}

\begin{figure}[!h]
\centering
\includegraphics[width=1.0\columnwidth]{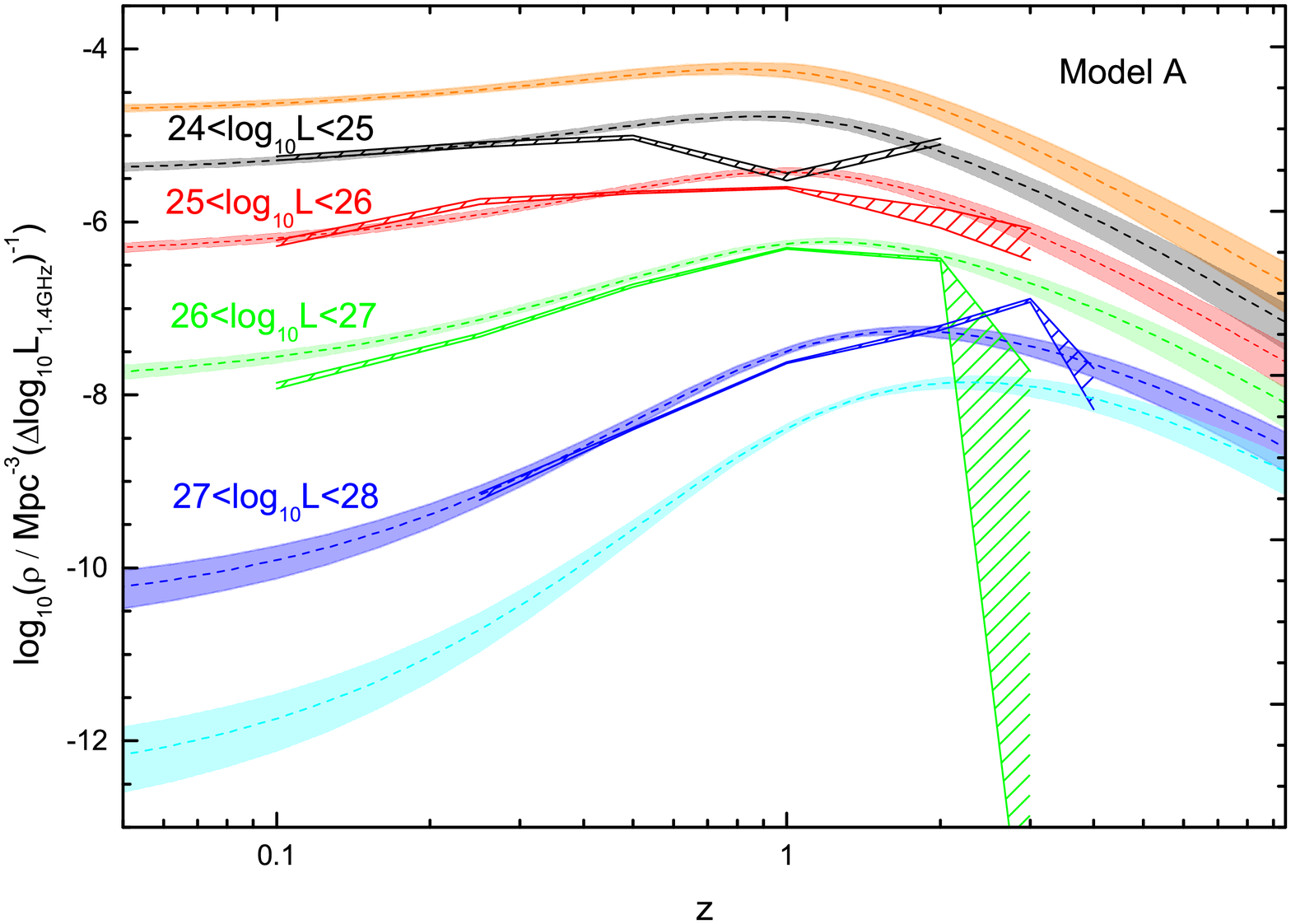}
\includegraphics[width=1.0\columnwidth]{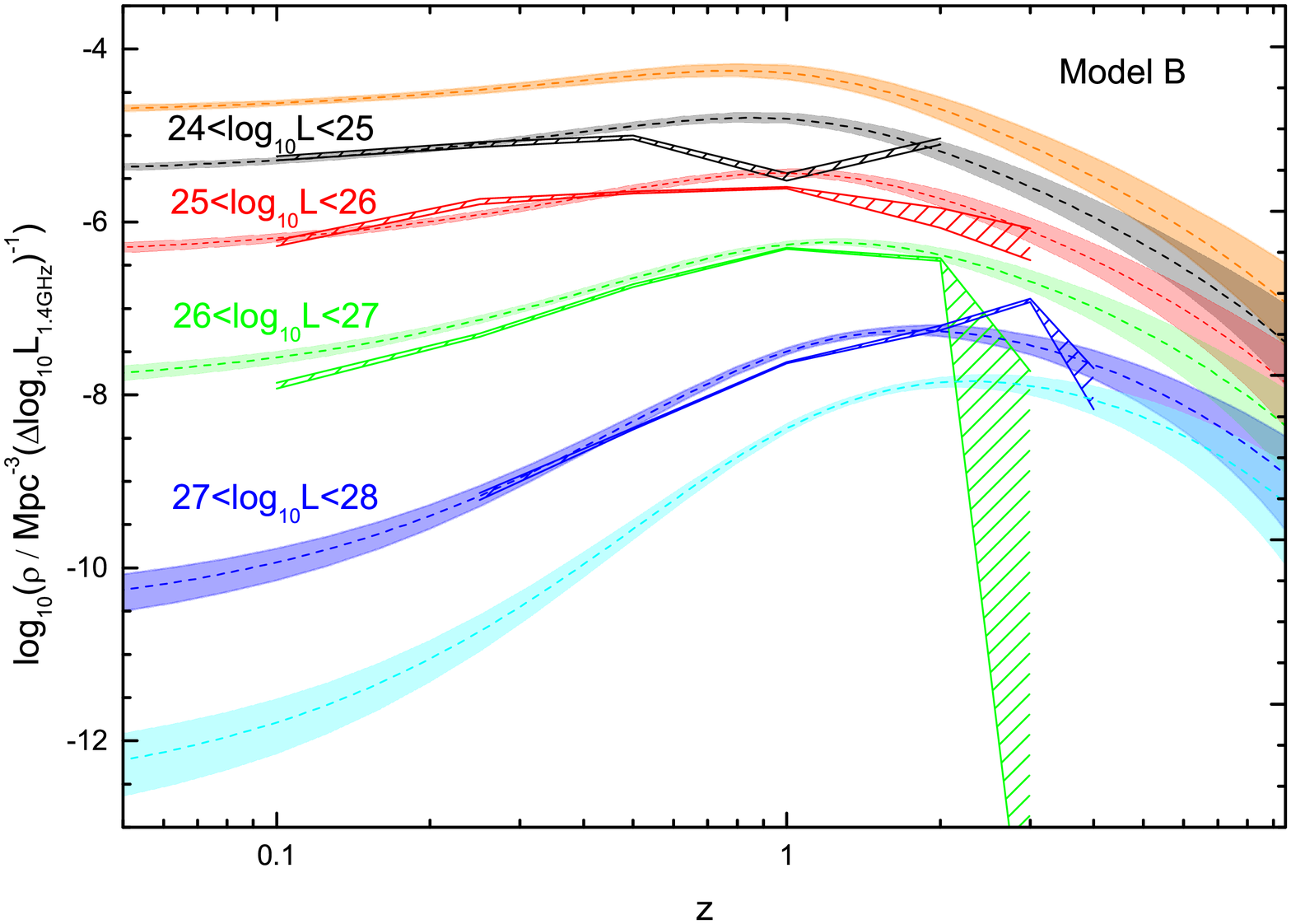}
\includegraphics[width=1.0\columnwidth]{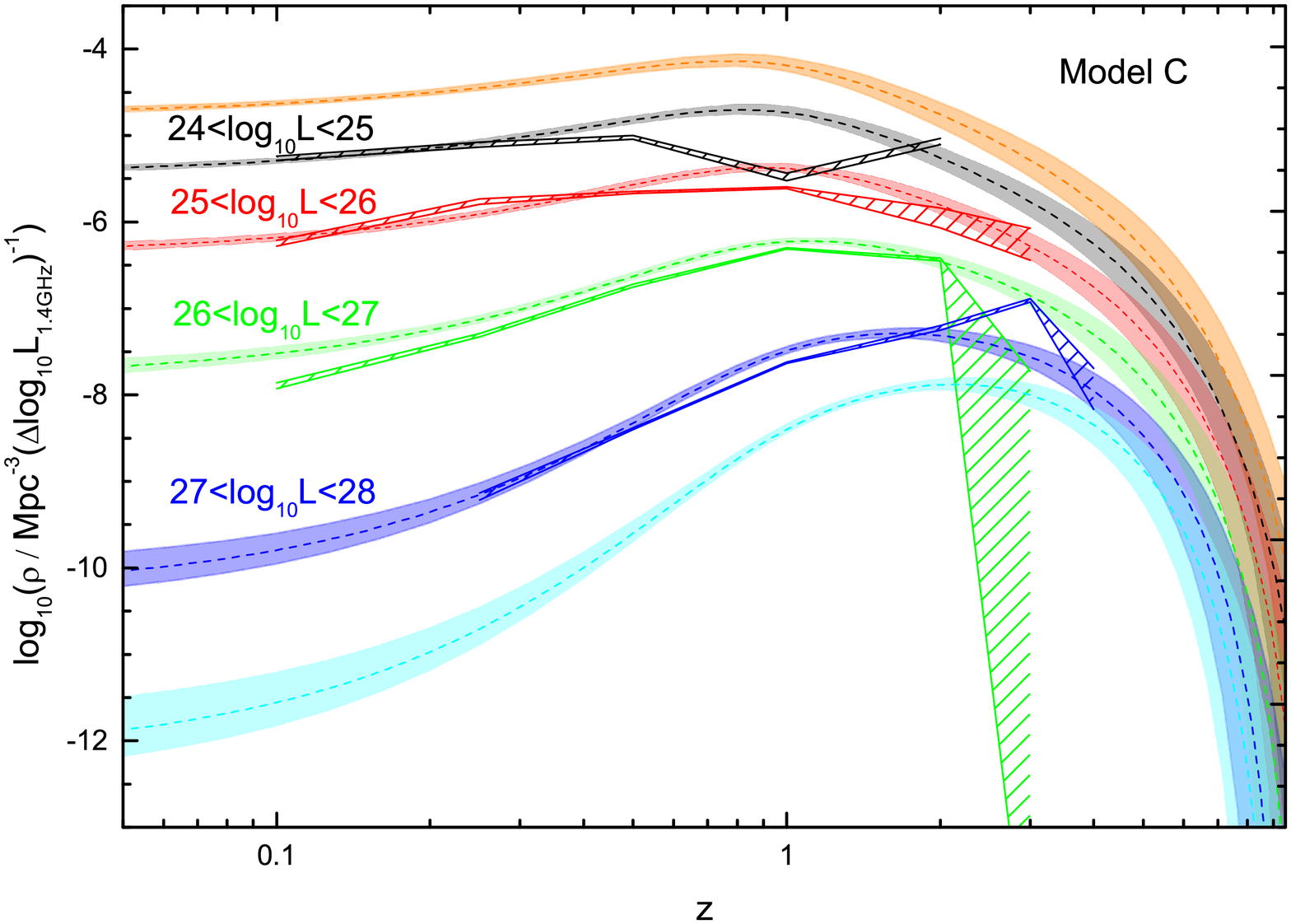}
\caption{Space densities as a function of redshift for Models A, B, and C, having been converted to the same frequency (1.4 GHz) used by R15. The orange, black, red, green, blue and cyan dashed lines show the RLFs at $\log_{10}L_{1.4\mathrm{GHz}}$=23.5, 24.5, 25.5, 26.5, 27.5 and 28.0 respectively. The light shaded areas take into account the 1 $\sigma$ error bands. The black, red, green and blue solid lines with shadows show the result of R15.}
\label{RLF-z}
\end{figure}

\begin{figure*}[!htb]
\centering
\includegraphics[height=5.1cm,width=5.89cm]{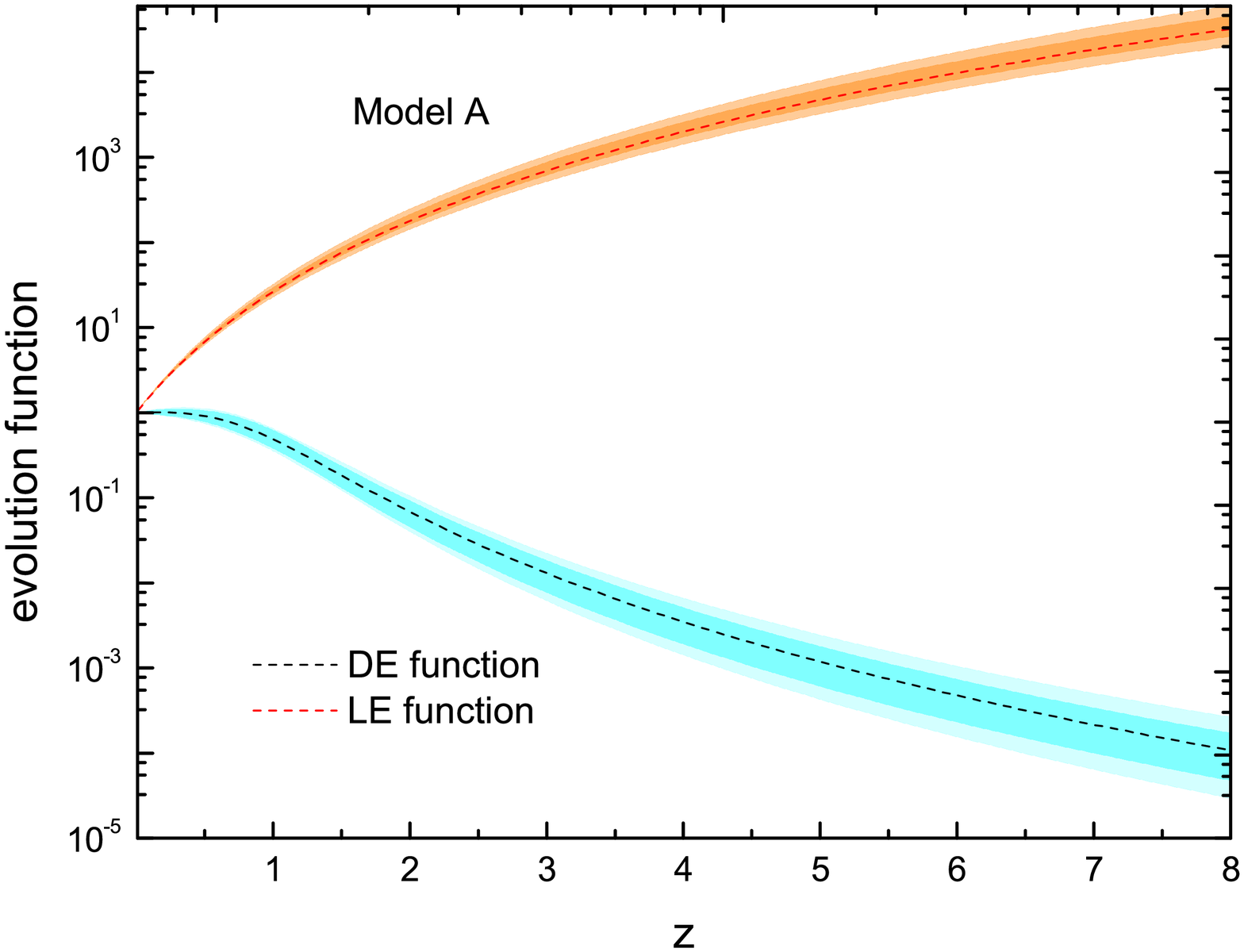}
\includegraphics[height=5.1cm,width=5.74cm]{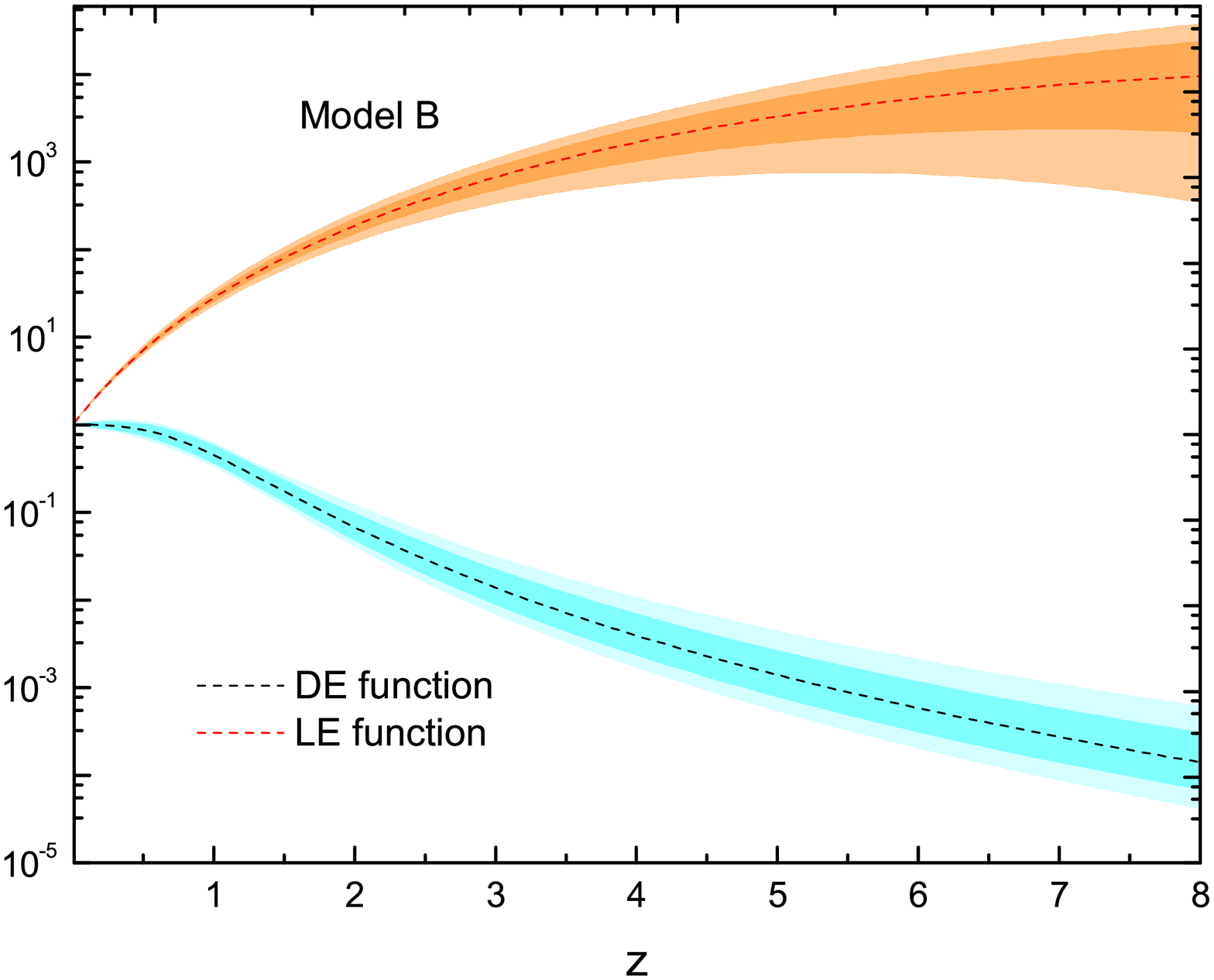}
\includegraphics[height=5.1cm,width=5.74cm]{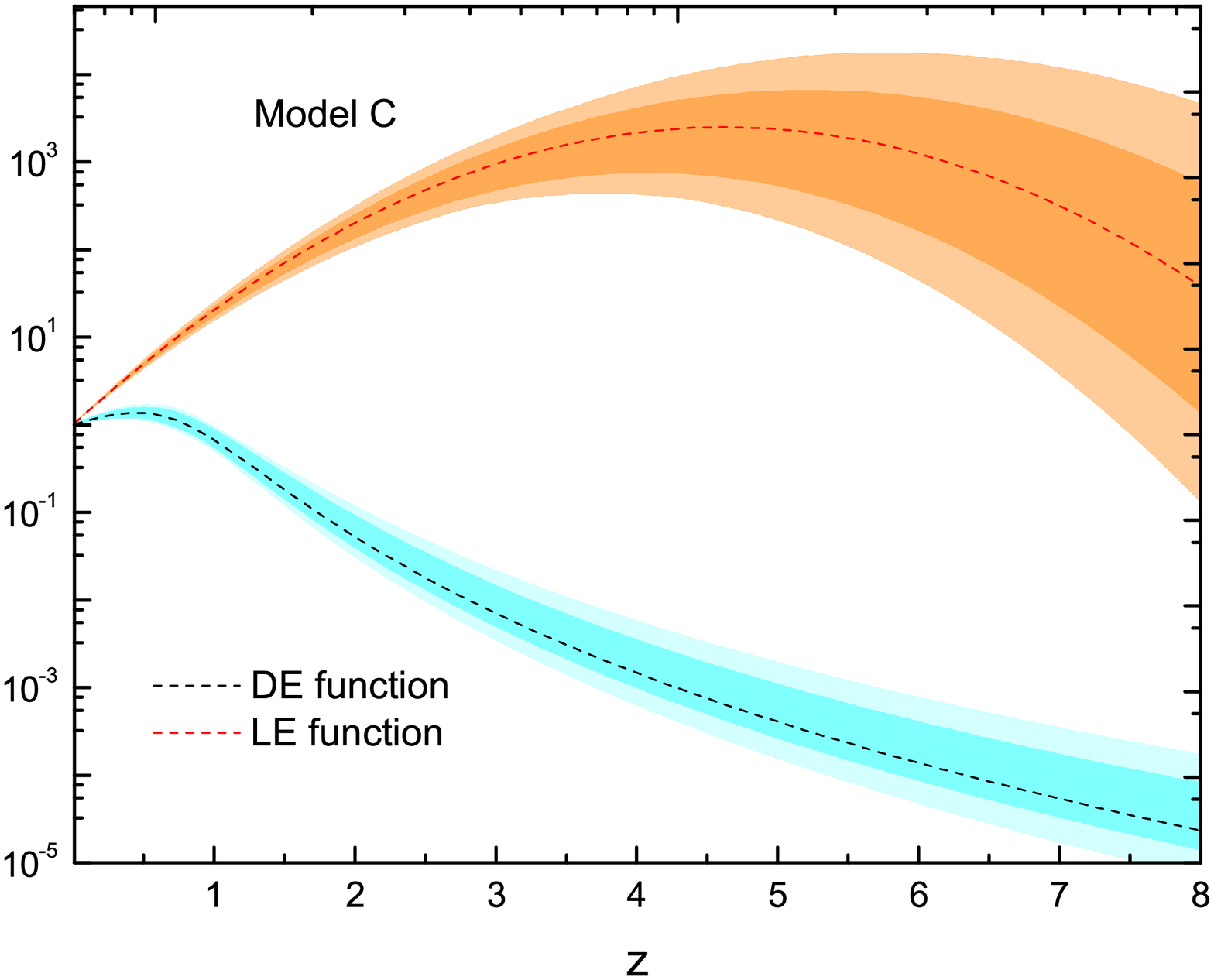}
\caption{DE and LE functions for models A, B and C. The black and red dashed lines represent the model DE and LE functions, respectively. The light shaded areas take into account the 1 $\sigma$ and 2 $\sigma$ error bands.}
\label{DELE}
\end{figure*}

\begin{figure}[!htb]
\centering
\includegraphics[width=1.0\columnwidth]{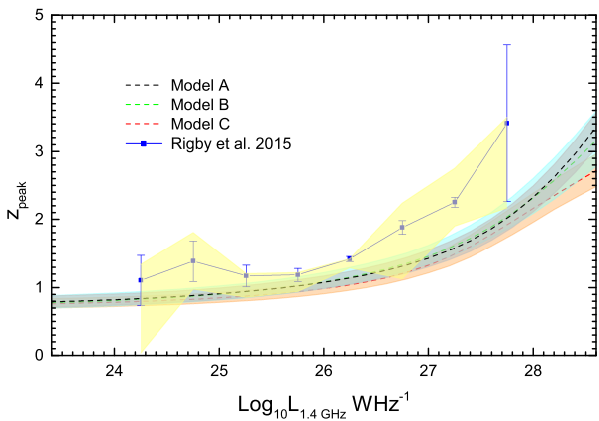}
\caption{Variation in the redshift of the peak space density with radio luminosity. The black, green, and red dashed lines represent Model A, B and C respectively. The light shaded areas take into account the 1$\sigma$ error bands. The blue filled squares with error bars show the result of R15. The light yellow shaded region represents the range in their results found from varying the input parameters used in the RLF grid modelling.}
\label{zpeak}
\end{figure}

\begin{figure*}[!htb]
\centering
\includegraphics[width=2.2\columnwidth]{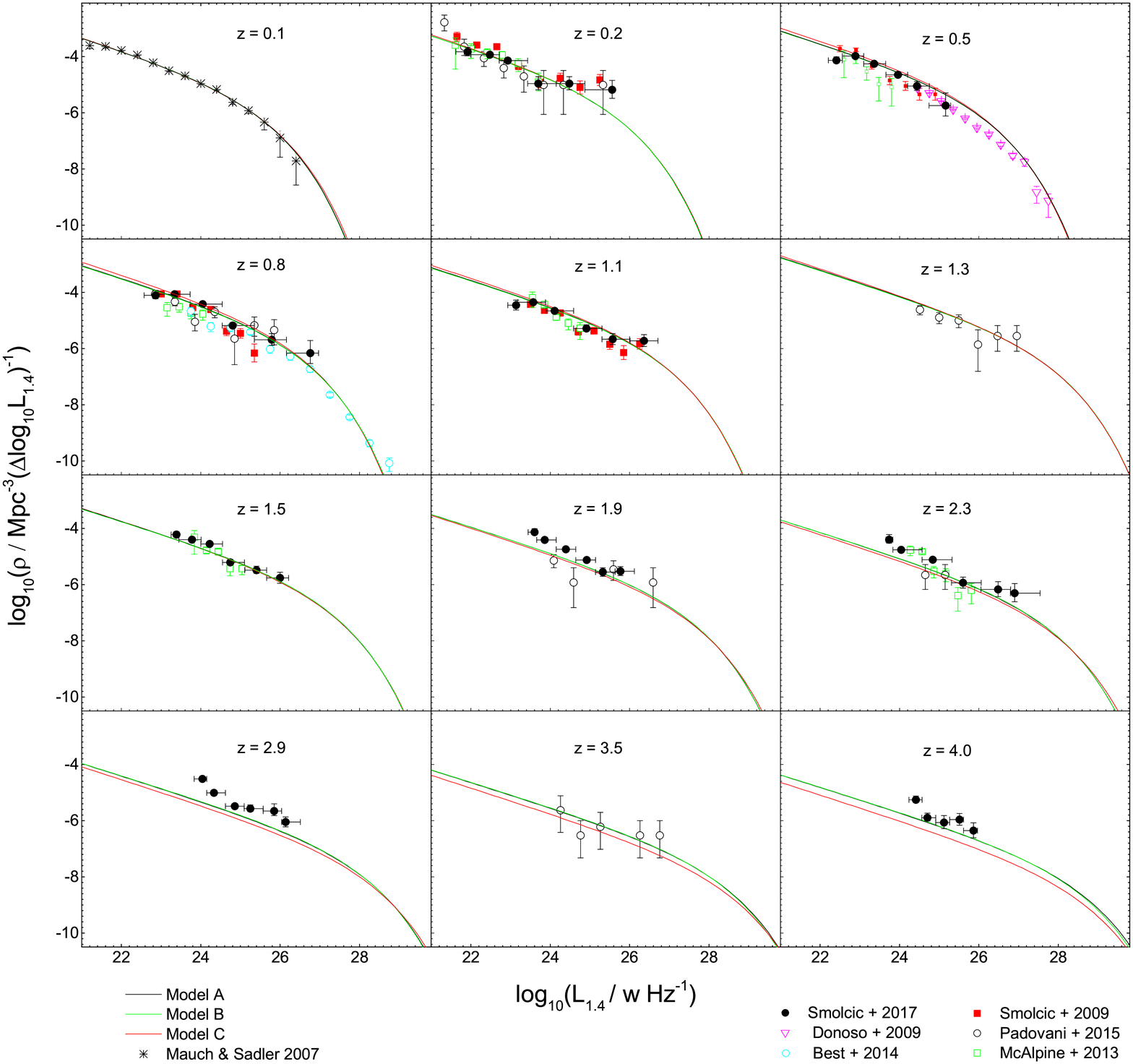}
\caption{Model RLFs (have been converted to 1.4 GHz by assuming a typical spectral index of 0.75) at various redshifts compared with the observational estimates specified in the inset. The RLFs for Model A, B and C are represented by black, green and red solid lines, respectively. }
\label{fig_comp}
\end{figure*}


\section[]{results}
\subsection{Model Parameters and RLFs}
Following Paper I, we use the public Fortran code ``CosmoMC'', which is based on a Markov chain Monte Carlo (MCMC) sampling algorithm \citep{2002PhRvD..66j3511L}, to estimate the best-fit parameters of our models. The best-fit parameters and their $1 \sigma$ error are reported in Table 1. The CosmoMC program also gives the one-dimensional (1D) probability distributions (left panels in Figure \ref{fig_mcmc}) and two-dimensional (2D) confidence contours (at 1$\sigma$ and 2$\sigma$ confidence levels, right panels in Figure \ref{fig_mcmc}) of the model parameters \citep[e.g.,][]{2013ApJ...765..122Y}. In Figure \ref{fig_mcmc}, we show these results for models A, B, and C in the upper, middle, and lower panels, respectively. Except for the parameter $k_2$ of model B, all the other parameters for our three models are well constrained. For the 2D confidence contours of the parameters, we only show the combinations with relatively large correlations \citep[e.g.,][]{2011ApJ...735..120Y}. Neither of contours for the parameters $\mu$ and $\sigma$ appears here, indicating that they are weakly correlated with the other parameters. This in turn confirms the rationality of previous assumptions that the spectral index distribution is independent of redshift and luminosity.

In Figure \ref{RLF-L}, we show the 408 MHz RLF yielded by model A at $z=0.1$, $0.5$, $1$, $2$, and $3$. Figure \ref{RLF-z} presents the space density changing with redshift for our three models at various luminosities. These are in broad agreement with the results of Paper I (see their Figure 2 and 3) for low redshift ($z<1.0$). This work updates the RLFs for high redshift (z$>$1) based on additional constraints provided by radio counts and the local RLF.

\subsection[]{DE and LE}
\label{DELE}
The key point of our mixture evolution scenario is that the DE and LE jointly dominate the evolution of radio-loud AGNs. In Figure \ref{DELE}, we show the DE and LE functions for Models A, B, and C (in left, middle, and right panels, respectively). The DE function slowly increases as $(1+z)^{0.3 \thicksim 1.3}$ out to $z \thicksim 0.8$, and then rapidly decreases as $(1+z)^{-6.8 \thicksim -5.7}$. The LE function rapidly increase as $(1+z)^{4.8}$ out to a higher redshift (at least $z>3.5$). Model C permits the possibility of negative LE at high redshift. However, from the right panel of Figure \ref{DELE} we can see that the uncertainty increases considerably when $z>3.5$ (it is the maximum redshift of our sample), reflecting the fact that the redshifts in our sample are not high enough to constrain a possible LE peak (or a flattening). In other words, if the LE has a peak redshift, it must be larger than 3.5.

It is also straightforward to demonstrate that the LE (Equation \ref{LEB}) of model B peaks at $\sqrt{k_1k_2^2+\frac{1}{4}}-\frac{1}{2}$. The parameter $k_2$ is very important to constrain the peak. However, from the middle panel of Figure \ref{fig_mcmc}, the sharp edge in probability distribution of the parameter $k_2$ for model B indicates that $k_2$ is poorly constrained. This further confirms the fact that the redshifts in our sample are not high enough to constrain a possible LE peak (or a flattening).

\subsection{Luminosity-dependent evolution}
As discussed in Paper I, the luminosity-dependent evolution is a natural consequence of our mixture evolution scenario. Indeed, in Figure \ref{RLF-z}, the predicted variation of space density with redshift clearly depends on radio luminosity. Moreover, the amount of space density change from redshift zero to the maximum space density is also a strong function of radio luminosity. The change is more than a factor of 100 at high luminosities, but it is less than a factor of 10 for low radio luminosities. This is similar to what has been observed for X-ray and optically selected QSOs \citep[e.g.,][]{2005A&A...441..417H,2007ApJ...654..731H}. In Figure \ref{RLF-z}, we also show the result of R15. For comparison, our RLFs have been converted to 1.4 GHz. It seems that our models are consistent with their result. It is noted that for the luminosity bin $24<\log_{10}L<25$, the RLF of R15 presents a concavity at $z\thicksim1$. This could not be a real drop of space density. It can be statistical error caused by inappropriately dividing redshift bins. As discussed by \citet{2013Ap&SS.345..305Y}, the estimated result of the binning method relies on how to chose the start point and width of each bin.

\subsection[]{The Redshift Cutoff}

It was already becoming apparent in the late 1960s that rapid increase in the space density of quasars with redshift did not continue beyond $z\thicksim 2.5$, and had to decline at higher redshifts. This high-redshift decline, regardless of its magnitude, has come to be known as the `redshift cutoff' \citep[e.g.,][]{1999ASPC..156..163S,2000MNRAS.319..121J}. In paper I, we also discussed the existence of a redshift cutoff for steep-spectrum radio sources. However, due to limitations of our methodology, we were unable to give compelling evidence for a redshift cutoff. In this work, thanks to constraints from radio counts and well determined local RLFs, we can give a more accurate analysis of the RLF based on a trivariate estimator. From Figure \ref{RLF-z}, it is clear that a redshift cut-off is present for all three models. The main difference among the three models is whether the cut-off is sharp or shallow.

The LE function of model A does not present a high-redshift decline, but this does not prevent a redshift cutoff to be occurring for model A. This indicates that whether a redshift cutoff occurs or not mainly depends on DE. Nevertheless, LE can decide whether a redshift cutoff is sharp or shallow. For example, models B and C have the same DE functions, but the LE function of model C presents a more notable high-redshift decline than that of model B. Consequently, the redshift cutoff for model C is clearly sharper than that for model B.

\citet{1990MNRAS.247...19D} claimed the first evidence for a redshift cutoff in the steep-spectrum radio sources. However, their samples were incomplete in redshift information, and the
results were limited by the accuracy of the photometric redshifts. In the subsequent decades, the redshift cutoff problem for the steep-spectrum radio sources has always been controversial \citep[e.g.][]{2001MNRAS.327..907J,2007MNRAS.375.1349C}. This situation was significantly improved when \citet{2011MNRAS.416.1900R} found a clear (at $>3\sigma$ significance) decline in the number density in steep-spectrum radio sources. Nevertheless, whether such decline is sharp or shallow is still an open question. We believe the main issue here is that current radio surveys still lack enough high-redshift sources so the LE of radio AGNs cannot be well constrained.

\subsection[]{Luminosity-dependent $z_{peak}$}

Figure \ref{zpeak} shows the variation in the redshift of the peak space density with radio luminosity. The black, green and red dashed lines represent Models A, B, and C respectively, with the light shaded areas taking into account the 1$\sigma$ error bands. The blue filled squares with error bars show the result of R15. As they used a grid-based modeling method (for details, see their paper), the range in results found from varying the input parameters was also presented by the light yellow shaded region. This shaded region partly overlaps with our model prediction, indicating our measurement is in broad agrement with their result. Nevertheless, it looks that the peak redshifts predicted by our models are systematically lower than that of R15. Such a difference can be explained as follows. First, the RLF directly obtained in this work is at 408 MHz but not at 1.4 GHz. To compare with the result of R15, we have converted the RLF to 1.4 GHz by assuming a power law ($L/\nu\propto \nu^{-\alpha}$, with $\alpha$ = 0.75). The accuracy of this conversion depends on the power-law assumption and also on the value of $\alpha$ adopted. Second, the spectral index distribution of the sample is important for determining the peak redshift. R15 assumed $\alpha=0.83+0.4\log_{10}(1+z)$ with an additional uncertainty of $\alpha\pm0.2$ at each redshift, while we use a redshift independent logarithm-normal function (Equation \ref{SID}) to model the spectral index distribution.

\subsection{Comparison with observational estimates}
In the past decades, there have been many observational estimates for the RLFs of radio-loud AGNs. Some representative works are reviewed as follows. \citet{2009MNRAS.392..617D} determined the 1.4 GHz RLF based on a catalog of 14,453 radio-loud AGNs with 1.4 GHz fluxes above 3.5 mJy in the redshift range of $0.4 < z < 0.8$. \citet{2009ApJ...696...24S} explored the cosmic evolution of radio luminous AGNs out to $z=1.3$ using a large sample of $\thicksim 600$ low-luminosity radio AGNs drawn from the VLA-COSMOS survey. \citet{2013MNRAS.436.1084M} presented a determination of the RLF for the VLA-VIDEO (the Visible and Infrared Survey Telescope for Astronomy Deep Extragalactic Observations) survey field using reliably identified sources (complete to a depth of 100 $\mu$Jy) with $\thicksim 10$ band photometric redshifts out to $z \thicksim 2.5$. Based on a combined sample of 211 radio-loud AGNs with $0.5 < z <1.0$, which are spectroscopically classified into jet-mode (radiatively inefficient) and radiative-mode (radiatively efficient) AGN classes, \citet{2014MNRAS.445..955B} presented the measurement of the RLF of jet-mode sources out to $z=1$. \citet{2015MNRAS.452.1263P} studied the RLF for radio-loud and quiet AGNs, respectively, based on the Extended Chandra Deep Field-South Very Large Array sample, which reached a flux density limit at 1.4 GHz of 32.5 $\mu$Jy at the field center and redshift $\thicksim 4$. Very recently, \citet{2017A&A...602A...6S} promoted the study of the RLF of radio-loud AGNs to a higher redshift of $z \thicksim 5$. In Figure \ref{fig_comp}, we compare our model RLFs (converted to 1.4 GHz) with these observational estimates specified in Figure \ref{fig_comp}. The RLFs for Models A, B, and C are represented by black, green and red solid lines, respectively. Overall, our models can well fit a wide variety of observational data. It is noted that the estimation of \citet{2009MNRAS.392..617D} (purple inverted triangles in Figure \ref{fig_comp}) is obviously lower than our model RLFs at $z=0.5$. This is because the Donoso et al. sample misses RGs falling outside the MegaZ Luminous Red Galaxy color selection criteria, as well as radio-loud quasars. On the whole, their estimate may be too low by $0.1-0.2$ dex \citep[see][]{2010MNRAS.404..532M}.

\subsection[]{Fitting the observed radio counts}

\begin{figure}[!htb]
\centering
\includegraphics[width=1.0\columnwidth]{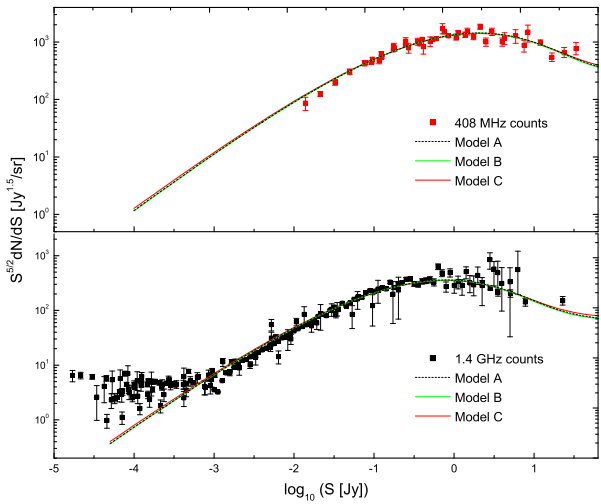}
\caption{Comparison of our best-fitting models with the observed radio counts at 408 MHz and 1.4 GHz (represented by red and black filled squares, respectively). The black dashed lines, green solid lines, and red solid lines show our best-fitting counts of model A, B and C, respectively.}
\label{counts}
\end{figure}

Another success of our model is that it can reproduce the observed radio counts. Figure \ref{counts} shows the comparison of our best-fitting models with the observed radio counts at 408 MHz and 1.4 GHz. The model counts at 1.4 GHz are extrapolated by assuming a typical spectral index of 0.75. Note that the bump at faint end of the 1.4 GHz observed counts is generally believed to be contributed by star-forming galaxies \citep[e.g.,][]{2010MNRAS.404..532M}.

\begin{figure}[!htb]
\centering
\includegraphics[width=1.0\columnwidth]{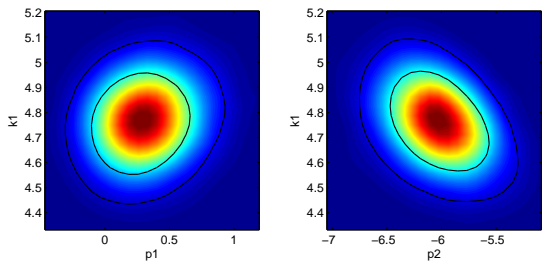}
\caption{1$\sigma$ and 2 $\sigma$ isoprobability contours of the likelihood estimated as functions of $k1$ and $p1$, and $k1$ and $p2$.}
\label{degeneracy}
\end{figure}

\section[]{Discussion}
\subsection{LADE versus LDLE}

Corresponding to the luminosity-dependent density evolution (LDDE) model, our mixture evolution model can also be referred to as the luminosity and density evolution (LADE) model \citep[also see][]{2010MNRAS.401.2531A}. The LDDE models successfully describe the X-ray LFs and $\gamma$-ray LFs \citep[e.g.,][]{2000A&A...353...25M,2012ApJ...751..108A,2013MNRAS.431..997Z}. We tried to use the same LDDE of \citet{2012ApJ...751..108A} to fit the steep-spectrum RLF, but we found that the fit is quite poor. A recent highly competitive model for the cosmological evolution of low-frequency radio sources was proposed by \citet{2010MNRAS.404..532M}. It is actually a luminosity-dependent luminosity evolution (LDLE) model, which successfully fitted a large amount of data on LFs of steep-spectrum sources, multi-frequency source counts and redshift distributions \citep[see][]{2017arXiv170405459B}. Our LADE model is also competitive in these aspects. Moreover, a notable advantage of the LADE model is that it does not need to assume in advance a luminosity-dependent $z_{peak}$ as in LDLE, and the luminosity-dependent evolution is just a natural consequence.

\subsection{Breaking the Evolution Degeneracy}
It is well known that the strong degeneracy between luminosity and density evolution \citep{2009ApJ...696...24S} means that we cannot distinguish specifically the contribution of DE and LE to the cosmic evolution of radio AGNs. Since the LE function of our model A only has one parameter, it is suitable for discussing this evolution degeneracy problem. Figure \ref{degeneracy} represents the 1 $\sigma$ and 2 $\sigma$ isoprobability contours of the likelihood estimated as functions of $k1$ and $p1$, and $k1$ and $p2$. The contours show near-circular shapes \citep[e.g.,][]{2005ApJ...632..169L}, reflecting the fact that the parameters k2, especially k1, are weakly correlated with p1. This means the degeneracy of DE and LE could be broken in our mixture evolution scenario.

\subsection{The physical meaning of LE and DE}
As presented in section \ref{DELE}, our models suggest a positive LE out to at least $z>3.5$, indicating that the radio-loud AGNs at high redshift are systematically brighter than that of today. Physically, this can be understood as follows. At higher redshifts, both the average density of the universe and the gas fraction are higher \citep{2014MNRAS.445..955B}, so that the radio lobes of AGNs remain more confined and adiabatic expansion losses are lower, leading to higher synchrotron luminosities \citep[e.g.,][]{1996MNRAS.283L..45B}. On the other hand, our models suggest that the DE peaks at $z\thicksim 0.8$ and then rapidly decreases, indicating that the radio-loud AGNs at high redshift are less numerous. This is broadly consistent with the result of \citet{2007ApJ...656..680J}, who found that the radio-loud fraction of quasars is a strong function of redshift, i.e. the radio-loud fraction decreases rapidly with increasing redshift.

\subsection{Do Low- and High-power Radio-loud AGNs evolve differently?}
There are three ingredients in our mixture evolution scenario. First, the shape of LF is independent of redshift and is always flat at low powers and steep at high powers. Second, the DE slowly increases out to a relatively low redshift and then rapidly decreases. Third, the LE increases out to a relatively high redshift. Once the above three conditions are met, a luminosity-dependent evolution, even a redshift cutoff will be a natural consequence. To explain the luminosity-dependent evolution of radio-loud AGNs, astronomers usually assume that the low- and high-power radio-loud AGNs evolve differently \citep[e.g.,][]{2001MNRAS.328..882W,2004MNRAS.352..909C}, which has been the mantra in radio astronomy for many years \citep{2016A&ARv..24...13P}. However, according to our mixture evolution scenario, there appears to be no need for different evolution for the low- and high-power radio-loud AGNs. It seems that both populations experience the same combined evolution of DE and LE. As the RLF is always flat at low powers and steep at high powers, the low-power sources are more sensitive to DE, while the high-power sources are more sensitive to LE (see the Figure 5 of Paper I). Consequently, the inferred turnover redshift for low-power sources is lower than that of high-power sources, mimicking a luminosity-dependent evolution.

\section[]{Conclusions}

The main results of this work are as follows.

\begin{enumerate}
  \item Following Paper I, we further confirm that the cosmic evolution of steep-spectrum radio-loud AGNs can be reproduced by a simple combination of DE and LE. This mixture evolution scenario can naturally explain the luminosity-dependent evolution of radio-loud AGNs. Our models successfully fit a large amount of data on RLFs of steep-spectrum sources and multi-frequency source counts. The models indicate that the DE slowly increase as $(1+z)^{0.3 \thicksim 1.3}$ out to $z \thicksim 0.8$, and then rapidly decreases as $(1+z)^{-6.8 \thicksim -5.7}$, while the LE rapidly increase as $(1+z)^{4.8}$ out to a higher redshift (at least $z>3.5$).

  \item We find a high-redshift decline (i.e. redshift cutoff) in the number density of steep-spectrum radio sources, but we cannot conclude whether such a decline is sharp or shallow. We believe that whether a redshift cutoff occurs or not depends mainly on DE, while LE can decide its steepness. To differentiate the sharp and shallow decline, the key is to use higher-redshift (at least $z>3.5$) radio samples to constrain the possible peak (or flattening) in the LE function.

  \item According to our mixture evolution scenario, there appears to be no need for different evolution for the low- and high-power radio-loud AGNs. Both types of sources experience the same combined evolution of DE and LE. As the RLF is always flat at low powers and steep at high powers, the low-power sources are more sensitive to DE, while the high-power sources are more sensitive to LE. Consequently, the inferred turnover redshift for low-power sources is lower than that of high-power sources, mimicking a luminosity-dependent evolution.
\end{enumerate}

\acknowledgments

We are grateful to the referee for useful comments that improved this paper. We acknowledge the financial support from the National Natural Science Foundation of China 11603066, 11173054, 11133006, 11163006, 11573060, 11673062 the Policy Research Program of Chinese Academy of Sciences (KJCX2-YW-T24), the CAS ``Light of West China" Program, the Strategic Priority Research Program, and the Emergence of Cosmological Structures of the Chinese Academy of Sciences (grant No. XDB09000000). J.M. is supported by the Hundred-Talent Program of Chinese Academy of Sciences, the Key Research Program of Chinese Academy of Sciences (grant No. KJZD-EW-M06), and the Introducing Overseas Talent Plan of Yunnan Province. Z.Y. thanks Yuan Zhao for helpful comments in English writing. The authors gratefully acknowledge the computing time granted by the Yunnan Observatories, and provided on the facilities at the Yunnan Observatories Supercomputing Platform.

\listofchanges

\end{document}